\documentclass[review,12pt]{elsarticle}

\usepackage{amssymb}
\usepackage{amsthm}
\usepackage{amsmath}
\usepackage{lmodern}
\usepackage{mathrsfs}
\usepackage{adjustbox}
\usepackage{graphicx}
\usepackage{epstopdf}
\usepackage{float}
\usepackage{caption}
\usepackage{subcaption}
\usepackage{bm}
\usepackage{bbm}
\usepackage{mathrsfs}
\usepackage{cleveref}
\usepackage{soul}
\usepackage{accents}
\usepackage[utf8]{inputenc}
\usepackage{textcomp}
\usepackage{graphicx}
\usepackage{xcolor}
\usepackage{courier} 
\usepackage{listings} 
\usepackage{tabu} 
\usepackage{longtable}
\usepackage[margin=2cm]{geometry}
\usepackage{changepage} 
\usepackage[inkscapelatex=false]{svg}
\usepackage{url}
\biboptions{sort&compress} 

\journal{Journal of Power Sources}

\makeatletter
\def\@author#1{\g@addto@macro\elsauthors{\normalsize%
    \def\baselinestretch{1}%
    \upshape\authorsep#1\unskip\textsuperscript{%
      \ifx\@fnmark\@empty\else\unskip\sep\@fnmark\let\sep=,\fi
      \ifx\@corref\@empty\else\unskip\sep\@corref\let\sep=,\fi
      }%
    \def\authorsep{\unskip,\space}%
    \global\let\@fnmark\@empty
    \global\let\@corref\@empty  
    \global\let\sep\@empty}%
    \@eadauthor={#1}
}
\makeatother

\begin{document}
 
\setcitestyle{square}

\begin{frontmatter}



\title{On the role of crack electrolyte wetting in the degradation and performance of battery active particles}


\author[OX]{Sebastian Luza-Vega}

\author[TU]{Ying Zhao}

\author[OX]{Emilio Mart\'{\i}nez-Pa\~neda\corref{cor1}}
\ead{emilio.martinez-paneda@eng.ox.ac.uk}

\address[OX]{Department of Engineering Science, University of Oxford, Oxford OX1 3PJ, UK}

\address[TU]{School of Aerospace Engineering and Applied Mechanics, Tongji University, Shanghai 200092, China}

\cortext[cor1]{Corresponding author.}

\begin{abstract} 
Cathode particle fracture is widely recognised as a major degradation mechanism in lithium-ion batteries, yet cracking also permits electrolyte wetting of newly exposed internal surfaces, modifying interfacial reaction pathways. The mechanistic role of electrolyte wetting in redistributing reactions within cracked particles remains unclear. Here, we isolate this effect through a controlled comparison between (i) a fully coupled electro-chemo-mechanical model resolving lithium concentration, electrostatic potential, and stress fields in both the active material and the electrolyte inside and outside cracks, and (ii) a single-particle chemo-mechanical model employing the conventional uniform flux assumption. The coupled model predicts strong spatial heterogeneity in interfacial reaction rates, with flux amplification approximately 8× relative to the imposed uniform flux at the crack tip. Reaction redistribution, and thus lithium flux, is governed predominantly by local solid-state lithium concentration and stress variations, while electrolyte potential gradients inside cracks remain secondary under the conditions considered. Uniform flux models can underpredict delivered capacity by 25\% at 1C-rate; this discrepancy increases at higher rates. They also underestimate tensile stresses throughout the delithiation process by 10\%, directly affecting crack driving conditions. These results demonstrate that neglecting crack–electrolyte coupling leads to systematic underestimation of both utilisation limits and fatigue-relevant stress histories.\\
\end{abstract}


\end{frontmatter}


\section{Introduction}
\label{Introduction}

Lithium-ion batteries play a central role in the transition toward low-carbon energy systems, particularly in the electrification of the transport sector. At present, the transportation sector is a major source of greenhouse gas emissions, motivating the widespread adoption of electric vehicles powered by lithium-ion batteries \cite{paper_transport}. Among current cathode technologies, Ni-rich layered oxides, such as $\mathrm{LiNiMnCoO_2}$ (NMC) and $\mathrm{LiNiCoAlO_2}$ (NCA), have attracted significant interest due to their high specific capacity. However, increasing the nickel content is accompanied by reduced thermal stability and accelerated capacity fading, which present major challenges for long-term battery performance \cite{Noh_2013, Han_2019}.

Cathode particle fracture is widely recognised as one of the main degradation mechanisms in lithium-ion batteries \cite{park_degradation_2019, Kondrakov_2017, Intergranular_Liu_2022}. In Ni-rich layered oxides, repeated and anisotropic volumetric changes associated with lithium insertion (lithiation) and extraction (delithiation) generate internal stresses that promote the formation and propagation of intergranular cracks within secondary particles. Once these cracks reach the particle surface, they enable electrolyte infiltration into the particle, triggering parasitic reactions that decompose the electrolyte, consume active material, and promote irreversible phase transitions, ultimately leading to capacity loss. In addition to these electrochemical effects, intergranular cracking can mechanically separate primary particles, degrading electronic connectivity within the secondary particle and increasing internal resistance \cite{Santos_2023, Ryu_2018_fading, Nam_2019, Liao_cracks_2023}. In extreme cases, this fragmentation can electrically isolate regions of the cathode particle, making parts of the active material electrochemically inaccessible \cite{Intergranular_Liu_2022, Makimura_isolated_2012}.

In recent years, numerous multi-physics models have been developed to gain a deeper understanding of battery degradation mechanisms \cite{Grazioli_2016_review, Zhao_2019_review, Pistori_2023_review}. Among these, single-particle models have been widely employed to investigate stress evolution, lithium diffusion, and fracture within active materials \cite{Klinsmann_delithiaion_single, Klinsmann_lithiaion_single, Allen_single_2021, Bai_single_2021, Taghikhani_single_2021, Singh_single_2022, Ai_2022_fatigue_crack, Crack_Shishvan_2023, Rezaei_single_2023, parks_direct_2023, Taghikhani_debonding, Asheri_single_2024}. The vast majority of these models do not explicitly consider the electrolyte and instead impose a uniform lithium flux over the particle boundary, implicitly assuming homogeneous interfacial kinetics \cite{Klinsmann_delithiaion_single, Klinsmann_lithiaion_single, Ai_2022_fatigue_crack, Bai_single_2021, Crack_Shishvan_2023}. In a related multiscale electro-chemo-mechanical study of sodium-ion batteries, Bhowmick \textit{et al.} \cite{bhowmick_2026} showed that such simplified uncoupled particle descriptions can misrepresent the effective capacity during cycling.

Experiments have shown that electrolyte wetting can play an important role in battery performance. When electrolyte penetrates into cracked cathode particles, it can shorten lithium diffusion pathways and increase the electrochemically active surface area available for charge transfer \cite{xia_2018_crack}. Ruess et al. \cite{Ruess_2020_influence} compared NMC particles cycled with liquid and solid electrolytes and found that cells with liquid electrolyte exhibited improved charge-transfer kinetics and a higher apparent lithium diffusion coefficient, resulting in enhanced Coulombic efficiency. Similar trends were reported by Trevisanello et al. \cite{Trevisanello_2021}, who compared single-crystal and polycrystalline NMC particles and showed that electrolyte infiltration in polycrystalline materials increases the apparent diffusion coefficient and reduces charge-transfer resistance. These observations highlight the presence of competing effects: while fracture generally accelerates degradation processes, electrolyte infiltration into newly formed crack surfaces can locally enhance lithium transport and interfacial reaction kinetics.

Some modelling studies have gone beyond the uniform-flux assumption by allowing spatial variations in reaction rate along the particle surface, while prescribing electrolyte conditions by imposing constant lithium-ion concentration and electrolyte potential \cite{Allen_single_2021, Taghikhani_single_2021}, or by assuming a uniform potential drop between the cathode active material and the electrolyte \cite{Singh_single_2022, Rezaei_single_2023, Asheri_single_2024}. More recently, efforts have been made to explicitly account for electrolyte penetration and wetting in cracked cathode particles. Chen et al. \cite{Chen_single_2025} developed a phase-field model capable of predicting crack propagation while permitting electrolyte wetting; however, the electrolyte itself is not explicitly modelled, and uniform electrolyte conditions are assumed inside and outside the cracks. Han et al. \cite{Han_2024} presented a coupled electro-chemo-mechanical model incorporating cohesive-zone damage to simulate crack propagation and electrolyte wetting. Nevertheless, the role of electrolyte wetting in controlling interfacial reaction redistribution within cracks remains unexplored. Yet experience from other occluded-electrolyte problems, such as corrosion, suggests that defect confinement can significantly alter local concentration and potential fields relative to the bulk \cite{Hageman_corrosion_evolution_2023, phase_field_corrosion_Cui_2023, Makuch_corrosion_2024}.

There is a need to understand how electrochemical reactions redistribute along cracked cathode surfaces, and how this redistribution affects mechanical evolution and lithium utilisation within individual particles. In this work, we address this gap by comparing two modelling frameworks that simulate a cracked cathode particle subjected to a full charge-discharge cycle, specifically designed to isolate the effect of interfacial reaction redistribution enabled by electrolyte wetting. The first is a fully coupled electro-chemo-mechanical model that explicitly resolves the evolution of concentration, potential, and stress fields in both the electrode and the electrolyte inside and outside cracks. The second is a chemo-mechanical model that assumes uniform electrolyte conditions. This controlled comparison allows us to isolate the role of electrolyte coupling in reaction redistribution, lithium utilisation, and stress evolution.

The remainder of the paper is organised as follows. Section 2 describes the modelling framework, including the geometry and governing equations. Section 3 details the numerical implementation and material properties. Section 4 presents and discusses the numerical results. Finally, Section 5 summarises the main findings.

\section{A fully-coupled electro-chemo-mechanical theory for electrode particles}
\label{Sec:Theory}

\begin{figure}[htp]
    \centering
    \includegraphics[width=0.9\linewidth]{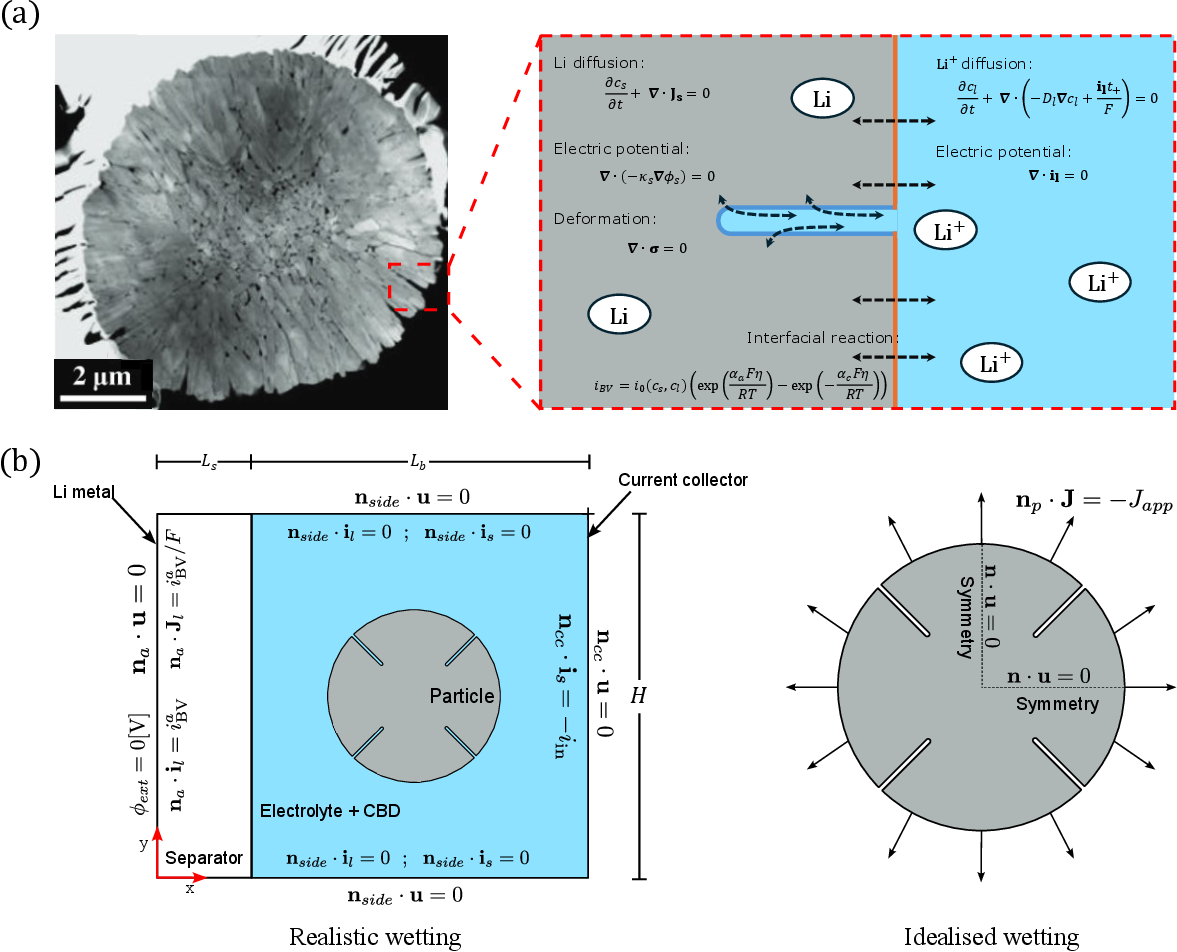}
    \caption{Physics and model geometry. (a): Overview of the problem under consideration: the behaviour of electrode particles exposed to an electrolyte. STEM image of a Ni-rich layered cathode particle (left, adapted from \cite{Park_2018_crack}) and a schematic illustrating the domain, governing equations and the crack electrolyte wetting process (right). In the schematic, the orange line denotes the uncracked surface and the blue line the cracked surface exposed to electrolyte infiltration. (b): Model geometries used in this work. In the realistic wetting model, a single cracked particle of radius 5 $\mu$m is embedded in a half-cell configuration. In the idealised wetting framework, the same particle geometry is considered, but without the influence of the electrolyte; a constant, uniform lithium flux is imposed along the particle boundary (including crack faces).}
    \label{fig: physic}
\end{figure}

As shown in Fig. \ref{fig: physic}(a), we study the evolution of the electrolyte both inside and outside cathode cracks, together with the response of the active material. Electrolyte wetting is explicitly simulated by solving the governing equations throughout the entire computational domain. As indicated by the arrows in Fig. \ref{fig: physic}(a), lithium insertion and extraction are allowed along the entire particle boundary, including the cracked surfaces. Below, we describe the governing equations of the fully coupled electro-chemo-mechanical model introduced here, along with a simpler reference model that does not include electrolyte wetting and instead uses the constant-flux approach commonly adopted in the literature. 

\subsection{Governing equations}

\subsubsection{Electrolyte and conductive matrix}
In the electrolyte domain, a binary liquid electrolyte is assumed and described using Newman’s concentrated solution theory. Electroneutrality is enforced, such that the cation and anion concentrations satisfy $c_l = c_+ = c_-$. The mass conservation equation for the electrolyte concentration reads,
\begin{equation}
    \frac{\partial c_l}{\partial t} + \boldsymbol{\nabla \cdot \mathrm{J}}_l = 0
    \label{eq: mass electrolyte} \, ,
\end{equation}

\noindent where $\boldsymbol{\mathrm{J}}_l$ is the lithium-ion flux within the electrolyte,  given by
\begin{equation}
    \boldsymbol{\mathrm{J}}_l = -D_l \boldsymbol{\nabla} c_l +  \frac{\boldsymbol{\mathrm{i_l}} t_+}{F},
    \label{eq: flux electrolyte}
\end{equation}
where $D_l$ denotes the electrolyte diffusion coefficient, $t_+$ the cation transference number, and $F$ the Faraday constant. Both diffusion and electromigration contributions to the transport of ionic species are accounted for. The electrolyte current density $\boldsymbol{\mathrm{i}}_l$ follows Newman’s concentrated solution theory \cite{fuller_1994}, 
\begin{equation}
    \boldsymbol{\mathrm{i}}_l = -\kappa_l \boldsymbol{\nabla} \phi_l + \frac{2 \kappa_l R_gT}{F} \left( 1 + \frac{\partial \ln(f)}{\partial \ln(c_l)} \right) (1 - t_+)\boldsymbol{\nabla} \ln(c_l),
    \label{eq: current density electrolyte}
\end{equation}
where $\kappa_l$ denotes the electrolyte conductivity, $\phi_l$ the electrolyte electric potential, $R_g$ the gas constant, $T$ the absolute temperature, and $f$ the mean molar activity.
The electrolyte electric potential is governed by charge conservation,
\begin{equation}
    \boldsymbol{\nabla \cdot \mathrm{i}}_l = 0.
    \label{eq: conservation charge electrolyte}
\end{equation}

The electrolyte domain provides a homogenised representation of a region containing liquid electrolyte and solid carbon binder, and thus provides an ionically and electronically conductive medium surrounding the active material. As such, deformation in the conductive matrix is captured by solving the mechanical equilibrium equation,
\begin{equation}
    \nabla \cdot (\widetilde{\boldsymbol{C}} ~ \nabla^{\mathrm{sym}} \textbf{u}) = 0,
\end{equation}
where $\textbf{u}$ is the displacement vector, $\nabla^{\mathrm{sym}}$ is the symmetric part of the spatial gradient, and $\widetilde{\boldsymbol{C}}$ is the homogenised stiffness tensor. The latter is obtained using a rule of mixtures, assuming a carbon binder volume fraction of 0.5 and typical values of Young's modulus and Poisson's ratio for carbon binder, 4 GPa and 0.3, respectively \cite{Xu_2017, xu_2019_heterogeneus}. 

Additionally, the carbon binder acts as an electronically conductive matrix that supports current transport. Under charge conservation, the electric potential in a solid phase satisfies
\begin{equation}
    \nabla \cdot \left(-\kappa_b \nabla \phi_s\right) = 0,
\end{equation}
where $\kappa_b$ is the electrical conductivity of the carbon binder, equal to 375 S\,m$^{-1}$ \cite{Boyce_2022}, and $\phi_s$ is the electric potential in the solid domain, which in this case corresponds to the carbon binder. 

\subsubsection{Active material}
Lithium concentration $c_s$ within the active material is governed by mass conservation,
\begin{equation}
    \frac{\partial c_s}{\partial t} + \boldsymbol{\nabla \cdot \mathrm{J}}_s = 0,
    \label{eq: mass_particle_1}
\end{equation}
with the lithium flux defined in terms of the chemical potential gradient, and the Li mobility coefficient $M(c_s)$ as:
\begin{equation}
    \boldsymbol{\mathrm{J}}_s = -c_s M(c_s)\boldsymbol{\nabla} \mu.
    \label{eq: flux_movility}
\end{equation}
The chemical potential of lithium in the active material incorporates mechanical effects,
\begin{equation}
    \mu = \mu_0 + R_gT \left( \ln \left(\frac{c_s}{c_{s_{max}}} \right) - \ln \left( 1 - \frac{c_s}{c_{s_{max}}} \right) \right) - \Omega(c_s) \sigma_h,
    \label{eq: chemical potential}
\end{equation}
where $\mu_0$ is the reference chemical potential, $\Omega$ the partial molar volume of lithium, $\sigma_h$ the hydrostatic stress, and $c_{s_{\max}}$ the maximum lithium concentration. The hydrostatic stress $\sigma_h =  \mathrm{tr}(\bm{\sigma})/3$ is defined positive in tension, thus tensile hydrostatic stresses lower the chemical potential and promote local lithium accumulation.

The Li mobility inside the host material is defined by
\begin{equation}
    M(c_s) = \frac{(c_{s_{max}} - c_s) D_s }{R_gT c_{s_{max}}},
    \label{eq: mobility}
\end{equation}
which, together with Eqs. (\ref{eq: chemical potential}) and (\ref{eq: flux_movility}), yields the expression for the lithium flux,
\begin{equation}
    \boldsymbol{\mathrm{J}}_s= -D_s \left(\boldsymbol{\nabla} c_s - \frac{\Omega c_s}{R_g T} \left( 1 - \frac{c_s}{c_{s_{max}}} \right) \boldsymbol{\nabla} \sigma_h - \frac{c_s \sigma_h}{R_g T} \left( 1 - \frac{c_s}{c_{s_{max}}} \right) \frac{d \Omega}{d c_s} \boldsymbol{\nabla} c_s  \right).
    \label{eq: flux particle}
\end{equation}
The mechanical deformation of the active material is modelled using linear elasticity. The stress field is governed by mechanical equilibrium which, in the absence of body forces, reads
\begin{equation}
    \boldsymbol{\nabla \cdot \sigma} = 0,
    \label{eq: mechanical equilibreium}
\end{equation}
where $\boldsymbol{\sigma}$ is the Cauchy stress tensor, related to the elastic strain $\boldsymbol{\varepsilon}_{\mathrm{el}}$ through Hooke’s law, $\boldsymbol{\sigma} = \boldsymbol{C}:\boldsymbol{\varepsilon}_{\mathrm{el}}$, with $\boldsymbol{C}$ being the fourth-order stiffness tensor. 
The total strain is computed using the small-deformation formulation,
\begin{equation}
    \boldsymbol{\varepsilon} = \frac{1}{2}((\nabla \boldsymbol{\mathrm{u}})^T + \nabla \boldsymbol{\mathrm{u}}),
    \label{eq: total strain}
\end{equation}
The total strain is decomposed into elastic and concentration-induced components, $\boldsymbol{\varepsilon} = \boldsymbol{\varepsilon}_{\mathrm{el}} + \boldsymbol{\varepsilon}_{ch}$. The concentration-induced strain is given by
\begin{equation}
    \boldsymbol{\varepsilon}_{ch} = \frac{1}{3} \left(\int_{c_{s_0}}^{c} \Omega (\xi) d\xi \right) \boldsymbol{\mathrm{I}},
    \label{eq: concentration strain}
\end{equation}
where $c_{s_0}$ is the initial lithium concentration, and $I$ is the second-order identity tensor.

In the active material, the electric current density follows Ohm's law,
\begin{equation}
    \boldsymbol{\mathrm{i}}_s = -\kappa_s \boldsymbol{\nabla} \phi_s,
    \label{eq: Ohms law electrode}
\end{equation}
where $\kappa_s$ denotes the electric conductivity in the active material, and $\phi_s$ refers here to the solid electrostatic potential in the active material domain. Due to charge conservation, the electrode current density follows,
\begin{equation}
    \boldsymbol{\nabla \cdot \mathrm{i}}_s = 0
    \label{eq: charge conservation electrode}
\end{equation}

\subsubsection{Electrolyte-particle interface}

At the interface between the active material and the electrolyte, the interfacial charge transfer reaction is described using Butler--Volmer kinetics,
\begin{equation}
    i_{\mathrm{BV}} = i_0 \left( \exp\left(\frac{\alpha_a F \eta}{R_gT}\right) - \exp\left(-\frac{\alpha_c F \eta}{R_gT}\right)\right),
    \label{eq: B-V equation}
\end{equation}
where $i_0$ is the exchange current density, $\alpha_a$ and $\alpha_c$ are the anodic and cathodic transfer coefficients, respectively, and $\eta$ is the overpotential. The exchange current density is assumed to depend on both the lithium concentration in the active material and the lithium-ion concentration in the electrolyte,
\begin{equation}
    i_0 = F k (c_{s_{max}} - c_s)^{\alpha_a} (c_s)^{\alpha_c} \left( \frac{c_l}{c_{l_{ref}}}\right)^{\alpha_a},
    \label{eq: exchange current density}
\end{equation}
with $k$ the reaction rate constants, and $c_{l_{ref}}$ the reference lithium-ion concentration in the electrolyte, typically taken as unity. 
The overpotential explicitly incorporates mechanical effects through the hydrostatic stress contribution,
\begin{equation}
    \eta = \phi_s - \phi_l - E_{eq} - \frac{\Omega \sigma_h}{F},
    \label{eq: overpotential}
\end{equation}
where $E_{eq}$ is the equilibrium potential. Positive overpotential $\eta$ corresponds to delithiation, i.e. lithium extraction from the particle.

\subsection{Idealised wetting}
The above-described coupled electro-chemo-mechanical model of active particles in liquid electrolytes is benchmarked against a reference chemo-mechanical particle model in which the electrolyte domain and interfacial electrochemical reactions are not explicitly resolved. Instead, the cathode particle is subjected to a prescribed, spatially uniform lithium flux at its surface, as is most often done in the literature to assess the role of cracks in electrode particle degradation. In this case, the governing equations are the same as those used for the solid electrode particle in the electro-chemo-mechanical electrolyte-particle model; Eq. (\ref{eq: mass_particle_1}) - (\ref{eq: flux particle}) are used to solve the lithium concentration, and Eq. (\ref{eq: mechanical equilibreium}) - (\ref{eq: concentration strain}) to solve the displacement field.

\section{Modelling details}
\label{Sec:FEM}

Two geometries are considered in this work: one for the electro-chemo-mechanical model, which will be referred to as realistic wetting (RW), and another for the chemo-mechanical model, referred to as idealised wetting (IW), as depicted in Fig. \ref{fig: physic}(b). The former (RW) represents the scenario where lithium diffusion in the electrolyte is resolved inside the crack, and the latter (IW) represents the prescribed homogeneous flux boundary condition at the crack surface. Here, the term realistic wetting is used only as a relative label to denote the electrolyte-resolved model considered in this work. It should not be interpreted as a fully realistic description of the electrolyte-active material interplay, since the present framework does not explicitly resolve all the physical mechanisms associated with electrolyte exposure and degradation of cathode particles.

A half-cell configuration is used in the realistic wetting case, consisting of a lithium-metal anode, a separator, a composite positive electrode, and a current collector. A plane strain condition is assumed. The anode and the current collector are incorporated via suitable boundary conditions, while the separator has dimensions $H = 20~\mu\mathrm{m}$ and $L_s = 5~\mu\mathrm{m}$. Within the composite electrode, the cathode particle is embedded in a homogenised positive-electrode domain comprising an electronically conductive carbon-binder phase and an ionically conductive electrolyte phase, with both width and height equal to $H = L_s = 20~\mu\mathrm{m}$. Accordingly, the particle surface is not electrically isolated: electronic current reaches the particle through the solid conductive network, while ionic current is transported through the electrolyte. To represent the homogenised domain, a porous medium formulation is adopted, in which the conductive matrix and liquid electrolyte are treated as a superimposed continuum \cite{Liu_2020_electrode}.

In both cases, we consider a cracked cathode particle with a radius of 5 $\mu$m. Since the focus of this work is on interfacial reaction redistribution rather than crack evolution, a simplified crack configuration is adopted in which one stationary surface crack is introduced in each quarter of the particle. This choice avoids mechanical interactions between neighbouring cracks. Each crack has a length of 1.77 $\mu$m and a width of 0.078 $\mu$m. These dimensions are representative of surface cracks observed experimentally in Ni-rich cathode particles \cite{xia_2018_crack, Ryu_2018_fading, Park_2018_crack, Nam_2019, Ryu_2020_crack, crack_lin_2020, capacity_ryu_2021, park_degradation_2019}, and are sufficiently large compared to the characteristic Debye length of the electrolyte double-layer to justify the assumption of electroneutrality within the cracks.

In the idealised wetting case, a single-particle geometry with the same dimensions, material properties, and mechanical description as the realistic wetting framework is used. Instead of computing the interfacial chemical reaction at the particle boundary, a constant and uniform lithium flux is prescribed along the particle surface, as it is commonly adopted in single-particle models in the literature. The idealised wetting model provides a baseline against which the impact of non-uniform electrochemical reactions can be quantified.

\subsection{Boundary and initial conditions}
\subsubsection{Realistic wetting}
Boundary conditions are defined with reference to the coordinate system shown in Fig. \ref{fig: physic}(b). A Li metal anode is assumed at $x = 0$, modelled as a zero-thickness boundary due to its high electronic conductivity. The electrolyte current density at the anode satisfies $\textbf{i}_l \cdot \textbf{n}_a = i_{\mathrm{BV}}^{a}$, with $i_{\mathrm{BV}}^{a}$ being the local interfacial current density, governed by Butler–Volmer kinetics,

\begin{gather}
        i_{\mathrm{BV}}^{a} = i_0^{a} \left( \exp\left(\frac{\alpha_a F \eta^{a}}{R_gT}\right) - \exp\left(-\frac{\alpha_c F \eta^{a}}{R_gT}\right)\right) \\
        i_0^{a} = i_{0_{ref}}^{a} \left( \frac{c_l}{c_{l_{ref}}} \right)^{\alpha_a} \\
        \eta^{a} = \phi_{ext} - \phi_{l} - E_{eq}^{a} \;,
\end{gather}
where $\phi_{ext}$ is an external electric potential equal to 0 V, and $E_{eq}^{a}$ is the equilibrium potential of the Li metal anode, also taken as 0 V. The flux at the anode is given by $\boldsymbol{\mathrm{J}}_{l} \cdot \boldsymbol{\mathrm{n}}_a = i_{\mathrm{BV}}^{a} / F$.

At the current collector boundary, $x = L_s + L_b$, a constant current density is imposed, $\textbf{i}_s \cdot \textbf{n}_{cc} = -i_{\mathrm{in}}$, corresponding to the current density needed to delithiate the particle in one hour (1C-rate). The applied current density is defined as
\begin{equation}
    i_{\mathrm{in}} = \frac{F \, c_{s_{\max}} \, V_p}{t_{1C} \, A_{cc}},
    \label{eq: imposed current density}
\end{equation}
where $V_p$ denotes the active material volume, $A_{cc}$ the current collector area, and $t_{1C} = 3600 \, \mathrm{s}$.

At the interface between the electrolyte and the active material, mass and charge transfer are coupled through the interfacial reaction rate $i_{\mathrm{BV}}$. The ionic and solid-phase lithium fluxes satisfy
\begin{equation}
    \boldsymbol{\mathrm{J}}_l \cdot \textbf{n}_p = -\frac{i_{\mathrm{BV}}}{F}, \qquad
    \boldsymbol{\mathrm{J}}_s \cdot \textbf{n}_p = \frac{i_{\mathrm{BV}}}{F},
\end{equation}
while the corresponding electronic current densities are given by
\begin{equation}
    \boldsymbol{\mathrm{i}}_s \cdot \textbf{n}_p = -i_{\mathrm{BV}}, \qquad
    \boldsymbol{\mathrm{i}}_l \cdot \textbf{n}_p = i_{\mathrm{BV}},
\end{equation}
where $i_{\mathrm{BV}}$ is defined by the Butler–Volmer equation, Eq.(\ref{eq: B-V equation}).

All remaining external boundaries at $y = 0$ and $y = H$ are assumed to be electronically and ionically insulating,
\begin{equation}
    \boldsymbol{\mathrm{i}}_s \cdot \textbf{n}_{\mathrm{side}} = 
    \boldsymbol{\mathrm{J}}_s \cdot \textbf{n}_{\mathrm{side}} =
    \boldsymbol{\mathrm{i}}_l \cdot \textbf{n}_{\mathrm{side}} =
    \boldsymbol{\mathrm{J}}_l \cdot \textbf{n}_{\mathrm{side}} = 0 .
\end{equation}

Mechanical boundary conditions are imposed as roller constraints along all external boundaries, such that the normal component of the displacement vanishes at the anode ($x=0$), the current collector ($x=L_s+L_b$), and the lateral boundaries ($y=0$ and $y=H$).

Initial conditions correspond to a uniform electrolyte concentration $c_{l_0}=1000$ mol/m$^{3}$ and a nearly fully lithiated particle with $c_{s_0}=0.95\,c_{s_{\max}}$.

\subsubsection{Idealised wetting}

For the idealised wetting case, the geometrical symmetry of the particle is exploited and only a quarter of the particle is modelled, as shown in Fig. \ref{fig: physic}(b). In terms of boundary conditions, an equivalent flux to the one experienced for the particle in the realistic wetting model is imposed along the whole boundary,
\begin{equation}
    J_{app} = \frac{\int_0^H i_{\mathrm{in}}/F \; \text{d}y}{\int_{\Gamma} \text{d}S},
\end{equation}
where $i_{\mathrm{in}}$ is computed using Eq. (\ref{eq: imposed current density}) and $\Gamma$ denotes the particle boundary.

Similar to the realistic case, the initial condition corresponds to a uniform lithium concentration equal to $c_{s_{0}} = 0.95 \; c_{s_{max}}$.

\subsection{Potential estimation}
The idealised wetting scenario, commonly adopted in the literature, is taken as the reference case; a single-particle domain without an explicit electrolyte in which a constant, uniform lithium flux is imposed along the particle boundary. Under this assumption, interfacial electrochemical reactions are not explicitly solved, and the electrostatic potential is not explicitly computed. In order to have a direct comparison of voltage–capacity behaviour between the idealised and realistic models, an approximate cell potential $V$ is computed for the former case.

Using the definition of the overpotential, Eq. (\ref{eq: overpotential}), the difference between the electrostatic potentials (i.e., voltage $V$) can be written as,
\begin{equation}
    V \approx\phi_s - \phi_l = \eta + E_{eq} + \frac{\Omega \sigma_h}{F}
    \label{eq: potential estimation 1}
\end{equation}

From the right-hand side in the equation above, the overpotential $\eta$ is the only quantity not directly available in the idealised wetting case. Assuming symmetric charge transfer ($\alpha_a = \alpha_c = 0.5$) and a negligible concentration drop, the overpotential is approximated as,

\begin{equation}
    \eta = \frac{2R_gT}{F} \mathrm{asinh}\left(\frac{i_{in}}{2i_0}\right),
    \label{eq: potential estimation 2}
\end{equation}
where $i_{in}$ is obtained from the imposed boundary flux. By assuming a constant electrolyte concentration $c_l=1000 \; \mathrm{mol~ m^{-3}}$, the exchange current density $i_0$ can be evaluated using Eq. (\ref{eq: exchange current density}), allowing the estimation of the cell potential in the idealised case.

\subsection{Material parameters}

\begin{figure}[htp]
    \centering
    \includegraphics[width=0.6\linewidth]{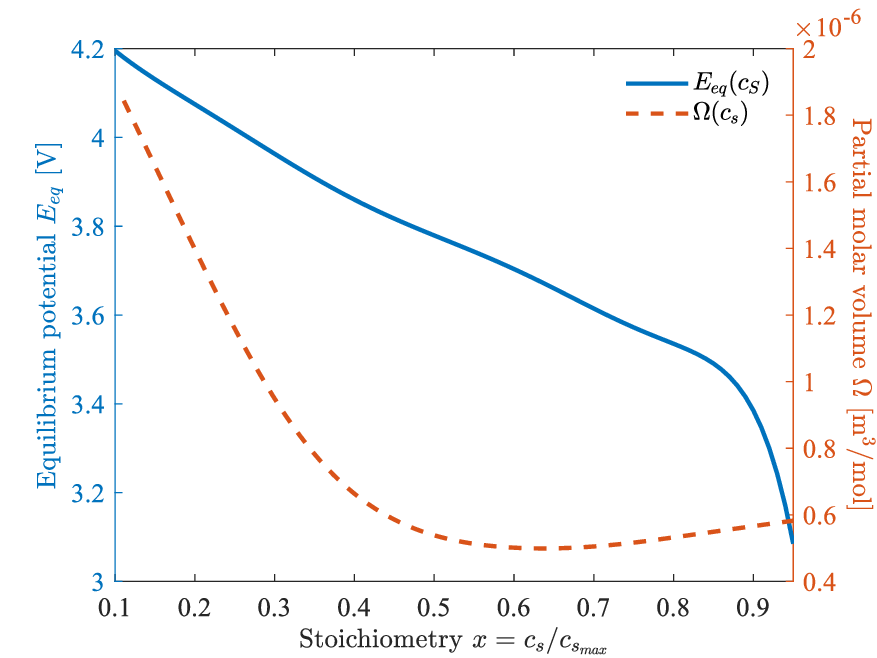}
    \caption{Variation of the equilibrium potential $E_{eq}$ and partial molar volume $\Omega$ as a function of the lithium stoichiometry and concentration ($x = c_s / c_{s_{\max}}$). Based on the experimental data reported in Refs. \cite{xu_2019_heterogeneus,yang_2024}.}
    \label{fig: potential validation and eq_omega}
\end{figure}

An NMC811 active particle is modelled as an isotropic continuum, without explicitly considering its primary particles, surrounded by an electrically conductive carbon binder matrix and an ionically conductive liquid electrolyte. The electrolyte is treated as a binary solution composed of an organic solvent and the lithium salt LiPF$_6$. The material properties used in the simulations are summarised in Table \ref{table material}, with transport coefficients adjusted using the Bruggeman relation $\kappa_{l,\mathrm{eff}} = \epsilon^{1.5} \kappa_l ~;~ D_{l,\mathrm{eff}}  = \epsilon^{1.5} D_l$, to account for a porosity $\epsilon = 0.5$. 

The equilibrium potential, $E_{eq}$, defined as the potential at which no net current flows, depends on the lithium stoichiometry $x = c_s / c_{s_{\max}}$. Similarly, the partial molar volume $\Omega$ depends on the lithium concentration within the particle and represents the contribution of one mole of lithium to the total volume of the active material. The variation of $E_{eq}$ and $\Omega$ with Li stoichiometry adopted in this work is given in Fig. \ref{fig: potential validation and eq_omega}, based on the experimental data reported in Refs. \cite{xu_2019_heterogeneus,yang_2024}.

\begin{table}[H]
    \scriptsize
    \begin{adjustbox}{width=\columnwidth,center}
    \begin{tabular}{|lcccc|}
    \hline
        Parameter & Symbol & Value & Unit & Reference\\
    \hline
        \multicolumn{5}{|c|}{Active material} \\
    \hline
        Diffusion coefficient & $D_s$ & $2.9\times10^{-15}$ & $\mathrm{m^{2} ~ s^{-1}}$ & \cite{Ashton_2021}\\
        Maximum concentration & $c_{s_{max}}$ & $4.93\times10^{4}$ & $\mathrm{mol ~ m^{-3}}$ & \cite{Gosh_2021}\\
        Young's modulus & $E_s$ & $150$ & $\mathrm{GPa}$ & \cite{parks_direct_2023}\\
        Poisson's ratio & $\nu_s$ & $0.3$ & & \cite{parks_direct_2023}\\
        Density & $\rho$ & $4780$ & $\mathrm{kg ~ m^{-3}}$ & \cite{Crack_Shishvan_2023}\\
        Equilibrium potential & $E_{eq}$ & $f(c_s)$ & V & \cite{xu_2019_heterogeneus}\\
        Partial molar volume & $\Omega$ & $f(c_s)$ & $\mathrm{m^{3} ~ mol^{-1}}$ & \cite{yang_2024}\\
    \hline
        \multicolumn{5}{|c|}{Electrolyte} \\
    \hline
        Ionic conductivity & $\kappa_l$ & $1.147$ & $\mathrm{S~m^{-1}}$ & \cite{valoean_2004}\\
        Diffusion coefficient & $D_l$ & $1\times10^{-10}$ & $\mathrm{m^{2} ~ s^{-1}}$ & \cite{valoean_2004}\\
        Transference number & $t_+$ & $0.363$ &  & \cite{valoean_2004}\\
        Activity & $\partial \ln(f)/ \partial \ln(c_l)$ & $0.43$ &  & \cite{valoean_2004}\\
    \hline
        \multicolumn{5}{|c|}{Interfacial reaction} \\
    \hline
        Reaction rate constant & $k$ & $2 \times 10^{-11}$ & $\mathrm{m~s^{-1}}$ & \cite{xu_2019_heterogeneus}\\
        Anodic transfer coefficient & $\alpha_a$ & $0.5$ &  & \cite{fuller_1994}\\
        Cathodic transfer coefficient & $\alpha_c$ & $0.5$ &  & \cite{fuller_1994}\\
    \hline
    \end{tabular}
    \end{adjustbox}
    \caption{Material properties used in the current study, categorised as a function of their domain of relevance (active particle, electrolyte and electrolyte-particle interface).}
    \label{table material}
\end{table}

\subsection{Numerical implementation}

The displacement field $\textbf{u}$, which describes particle deformation, is obtained from the mechanical equilibrium equation, Eq. (\ref{eq: mechanical equilibreium}). It is coupled to the solid-state lithium concentration $c_s$ through the chemical strain (Eq. (\ref{eq: concentration strain})), while the concentration field itself is determined by the mass conservation equation (Eq. (\ref{eq: mass_particle_1})). In turn, the solid-state concentration is influenced by mechanical stresses via the stress-dependent chemical potential (Eq. (\ref{eq: chemical potential})), thereby establishing a bidirectional coupling between the chemical and mechanical fields. The electric potential in the electrode $\phi_s$ is obtained from the charge conservation equation, Eq. (\ref{eq: charge conservation electrode}). Mechanical stresses also affect the interfacial charge transfer reaction by shifting the equilibrium potential. This effect is incorporated through the inclusion of the hydrostatic stress contribution in the overpotential, Eq. (\ref{eq: overpotential}).

Similarly, in the electrolyte domain, the lithium-ion concentration $c_l$ in the electrolyte is solved through the mass conservation equation (Eq. (\ref{eq: mass electrolyte})), while the electrostatic potential $\phi_l$ is calculated via the charge conservation equation, Eq. (\ref{eq: conservation charge electrolyte}). The electrolyte concentration and potential are coupled through the expression for the ionic current density $\textbf{i}_l$, Eq. (\ref{eq: current density electrolyte}).

The resulting system of non-linear equations is solved using the finite element method in COMSOL Multiphysics (v6.2)\footnote{The model will be made freely available at \url{https://mechmat.web.ox.ac.uk/codes} immediately after the article's acceptance}. Time integration is performed using an implicit backward differentiation formula (BDF), and all governing equations are solved in a fully coupled manner using the Multifrontal Massively Parallel Sparse (MUMPS) direct solver. After a mesh sensitivity study, a quadrilateral mesh with a total of 110,337 degrees of freedom is used for the idealised wetting framework. For the realistic wetting case, a similar quadrilateral mesh is used in the particle domain, while a triangular mesh is employed in the homogenised domain, giving a total of 387,043 degrees of freedom.

\section{Results and discussion}

To quantify the role of electrolyte wetting on both battery performance (assessed in Section \ref{sec: performance}) and particle integrity (Section \ref{sec: cracking}), we subject the realistic wetting and idealised wetting models to a complete charge-discharge cycle. During the charging (delithiation) stage, lithium is extracted from the particle until the minimum lithium concentration reaches a prescribed lower limit of $0.1\,c_{s_{\max}}$. At this point, the applied current density in the realistic wetting model and the imposed flux in the idealised wetting case are reversed, initiating the discharge (lithiation) process. The lithiation process continues until the maximum concentration in the particle equals the initial concentration $c_{s{_0}}$

We begin by examining how electrolyte wetting redistributes interfacial reaction rates along cracked surfaces and influences capacity utilisation. We first assess electrostatic potential within the electrolyte to evaluate potential transport limitations, before focusing on the resulting concentration and stress evolution within the cathode particle. The electrostatic potential distribution in the electrolyte is shown in Fig. \ref{fig: results 1}(a). Ionic transport from the current collector (right) toward the Li-metal anode (left) establishes a potential gradient, with the lowest electrolyte potential located near the anode where ions exit the positive electrode. Within the crack, the narrower and more confined transport pathway increases the local resistance to ion transport, leading to a notable elevation in electrolyte potential relative to the bulk. At the crack tip, the electrolyte potential reaches 1.1 mV, while the maximum potential in the bulk is 0.75 mV, corresponding to an increase of approximately 45\% inside the crack.

\label{Sec:Results3}
\subsection{Impact on performance}
\label{sec: performance}

\begin{figure}[H]
    \centering
    \includegraphics[width=0.9\linewidth]{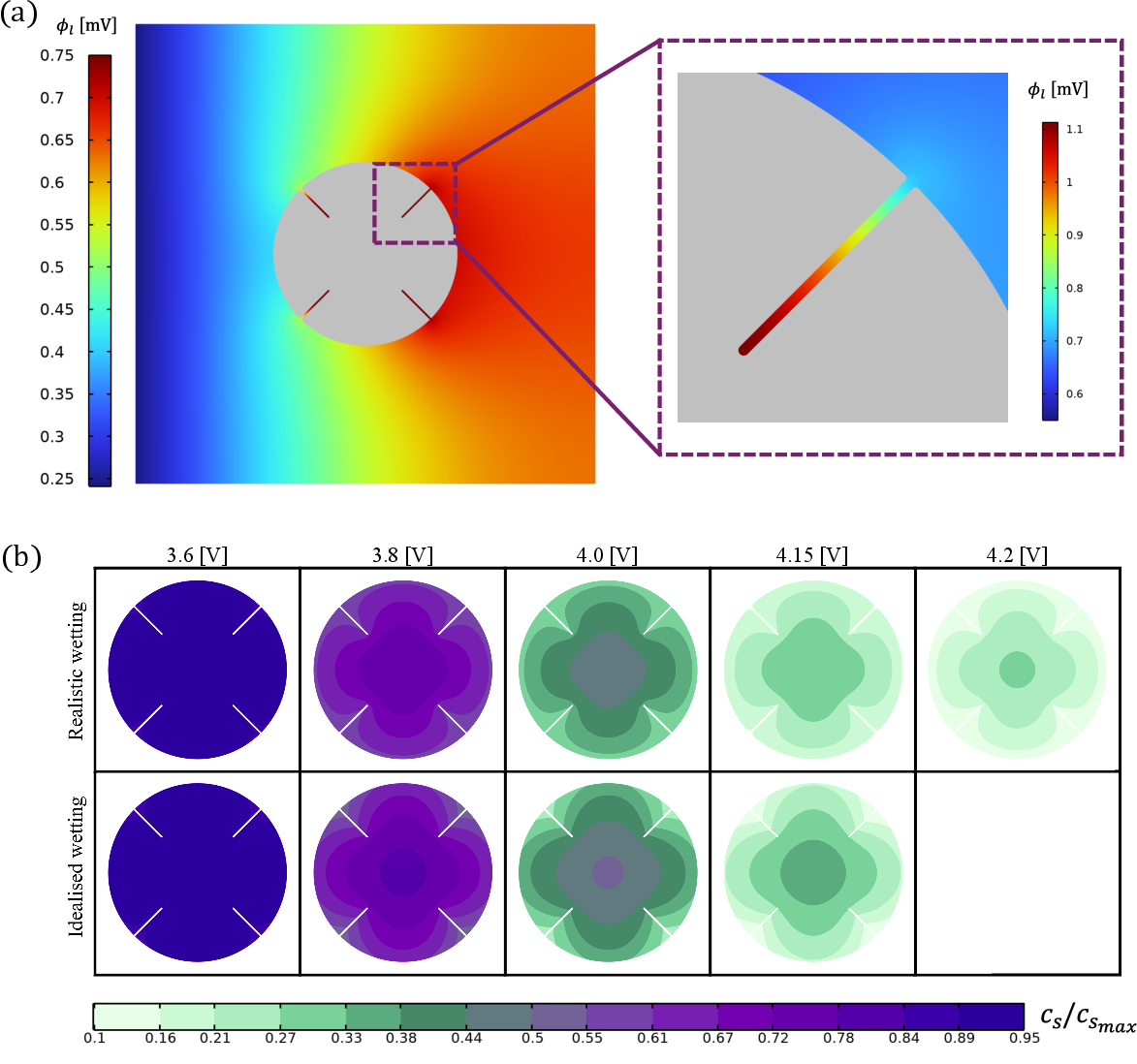}
    \caption{On electrolyte potential and Li content. (a): Electrolyte potential $\phi_l$ in the bulk electrolyte and inside the crack. The confined transport pathway elevates $\phi_l$ inside the crack, reaching 1.1 mV at the crack tip compared with 0.75 mV in the bulk electrolyte. (b): Normalised lithium concentration distribution within the cathode particle during delithiation for the realistic and idealised models. The idealised wetting case exhibits pronounced concentration heterogeneity, with lithium depletion at the crack opening and accumulation at the particle centre.}
    \label{fig: results 1}
\end{figure}

\begin{figure}[H]
    \centering
    \includegraphics[width=1.02\linewidth]{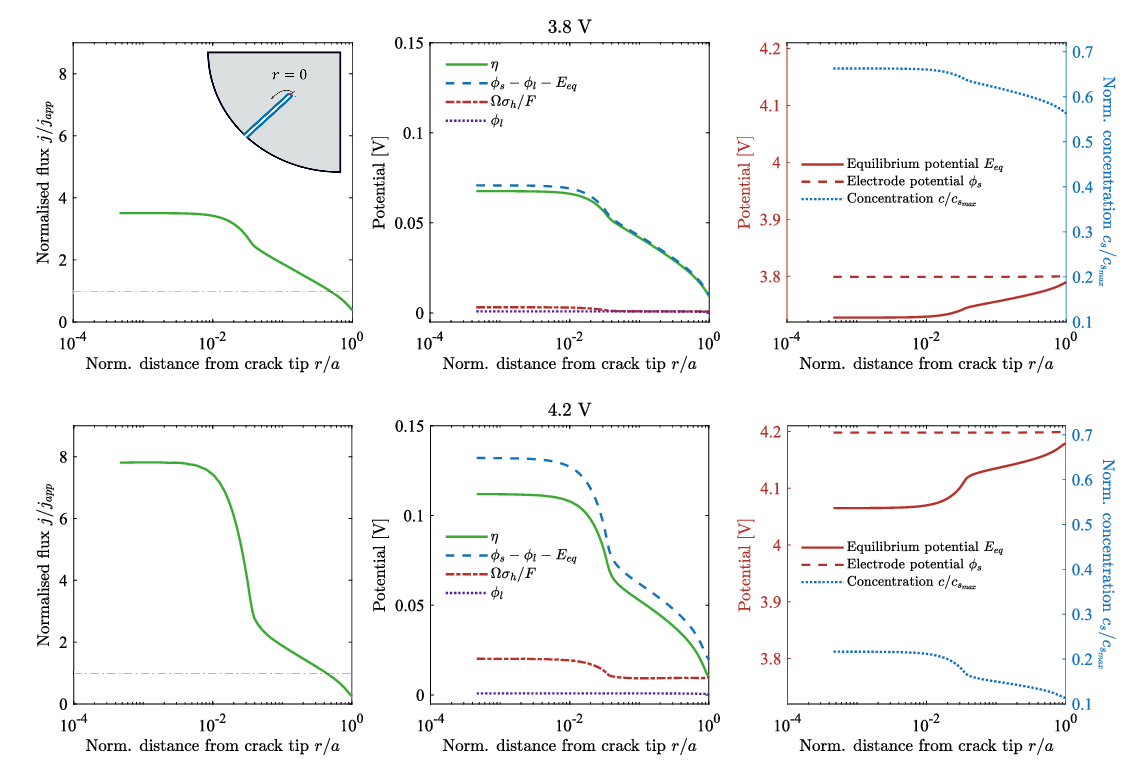}
    \caption{Spatial distribution of interfacial reaction rate along the cracked surface at two voltages: 3.8 V and 4.2 V. For each voltage, the columns (left to right) show: normalised lithium flux, overpotential and its contributions, and the electrode potential $\phi_s$, equilibrium potential $E_{eq}$, and normalised lithium concentration along the crack. Reaction-rate and overpotential amplification are observed near the crack tip, while variations in the electrolyte potential $\phi_l$ remain small. The electrode potential is nearly uniform along the crack, indicating that overpotential heterogeneity is governed primarily by variations in $E_{eq}$ arising from local lithium concentration gradients.}
    \label{fig: results 2}
\end{figure}

Regarding lithium concentration within the particle, the uniform-flux case exhibits a less homogeneous concentration, characterised by higher lithium accumulation at its centre and lower concentration at the crack openings, as shown in Fig. \ref{fig: results 1}(b). This depletion at the crack opening causes the particle to reach the minimal concentration threshold sooner than in the realistic wetting model. As a result, the idealised case finishes delithiation at a lower voltage (4.15 V) than the coupled framework (4.2 V), even though the same terminal conditions are imposed in both frameworks. This more homogeneous distribution within the particle originates from the spatial redistribution of lithium flux along the cracked surface in the realistic wetting model. This is shown in the left column of Fig. \ref{fig: results 2}, which depicts the normalised flux as a function of the distance ahead of the crack tip, showing a strong amplification of lithium flux at the crack tip during delithiation, reaching values almost 8 times larger than the uniform flux imposed in the idealised wetting case toward the end of the process, along with a reduction in flux at the crack opening. This high interfacial reaction rate at the crack tip depletes lithium more efficiently from the particle interior, leading to the lower concentration observed at the particle centre in the coupled framework, while the reduced flux at the crack opening promotes local lithium accumulation in that region.

The spatial distribution of lithium flux along the particle surface is governed by the local overpotential driving the interfacial electrochemical reaction. The centre column of Fig. \ref{fig: results 2} shows the overpotential distribution along the cracked surface during delithiation. At early stages, the overpotential is dominated by the electrostatic and equilibrium potential terms, i.e.\ $\eta \approx \phi_s - \phi_l - E_{\mathrm{eq}}$. As delithiation progresses, mechanical effects associated with stress-induced modifications of the chemical potential become more relevant, reducing the overall overpotential. The elevated overpotential observed near the crack tip is primarily associated with the electrode and equilibrium potential contributions. Despite the larger electrolyte potential gradients observed inside the cracks in Fig. \ref{fig: results 1}, variations in the electrolyte potential $\phi_l$ remain small compared with the magnitude of the overpotential. Consequently, the evolution of the overpotential along the crack is largely governed by the difference between the electrode potential and the local equilibrium potential.

The right column in Fig. \ref{fig: results 2} presents the electrode potential $\phi_s$ and the equilibrium potential $E_{eq}$ along the cracked surface. The electrode potential remains nearly uniform along the crack, demonstrating that the overpotential distribution is controlled by the equilibrium potential, which in turn depends on local lithium concentration. Although electrolyte wetting enables heterogeneous electrochemical reactions along cracked surfaces, the present results demonstrate that the resulting redistribution of interfacial reaction rates is governed primarily by lithium concentration gradients within the cathode particle. While both models, realistic and idealised, experience concentration gradients within the particle, imposing a uniform flux prevents these gradients from feeding back into the interfacial reaction rate. In contrast, variations in electrolyte potential inside the cracks remain small and do not significantly contribute to the observed reaction redistribution.

\begin{figure}[H]
    \centering
    \includegraphics[width=0.85\linewidth]{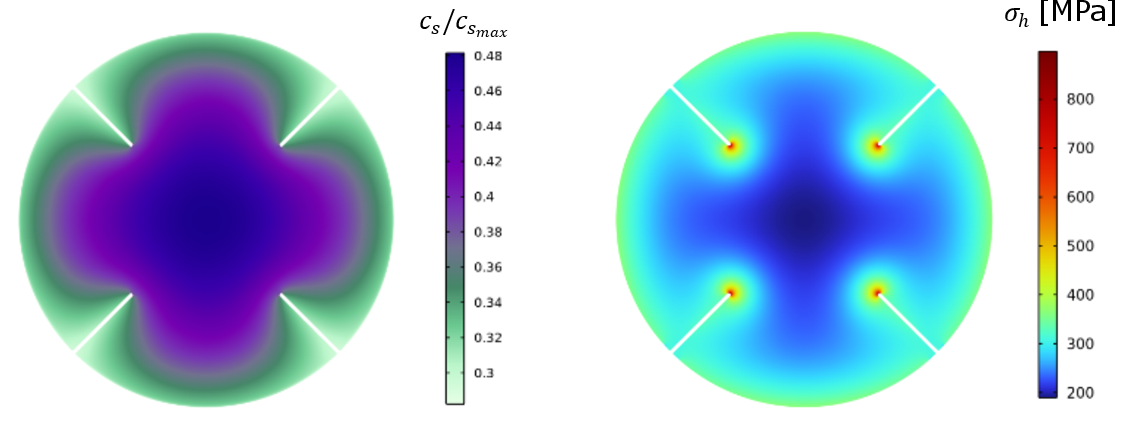}
    \caption{Normalised concentration and hydrostatic stress distribution inside the cathode particle at 4.0 V. The high stresses promote lithium accumulation through their effects on the chemical potential.}
    \label{fig: results 3}
\end{figure}

Despite the high lithium flux leaving the particle at the crack tip, this region also exhibits the highest lithium concentration along the particle boundary (see the left side of Fig. \ref{fig: results 3}). The mechanical effects on the chemical potential can explain this accumulation of lithium. As shown in Eq. (\ref{eq: chemical potential}), positive hydrostatic stresses reduce the chemical potential, promoting lithium accumulation, whereas compressive stresses increase the chemical potential and drive lithium away from the affected regions. As shown in Fig. \ref{fig: results 3} (right), the higher accumulation of lithium near the crack tip is the result of the high tensile hydrostatic stresses present in this region.

\begin{figure}[H]
    \centering
    \centering
    \subfloat[]{\includegraphics[width=0.47\linewidth]{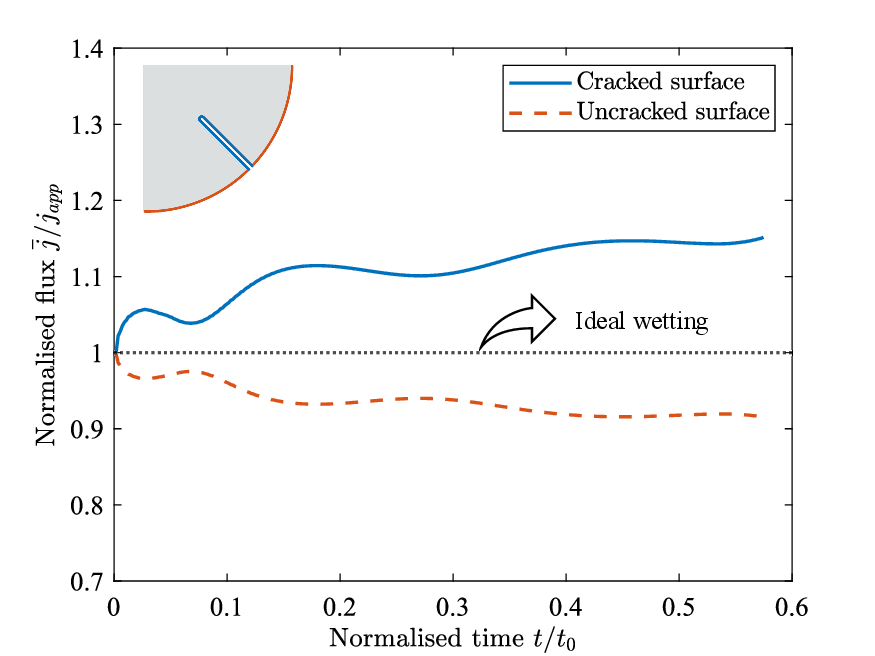}}
    \subfloat[]{\includegraphics[width=0.47\linewidth]{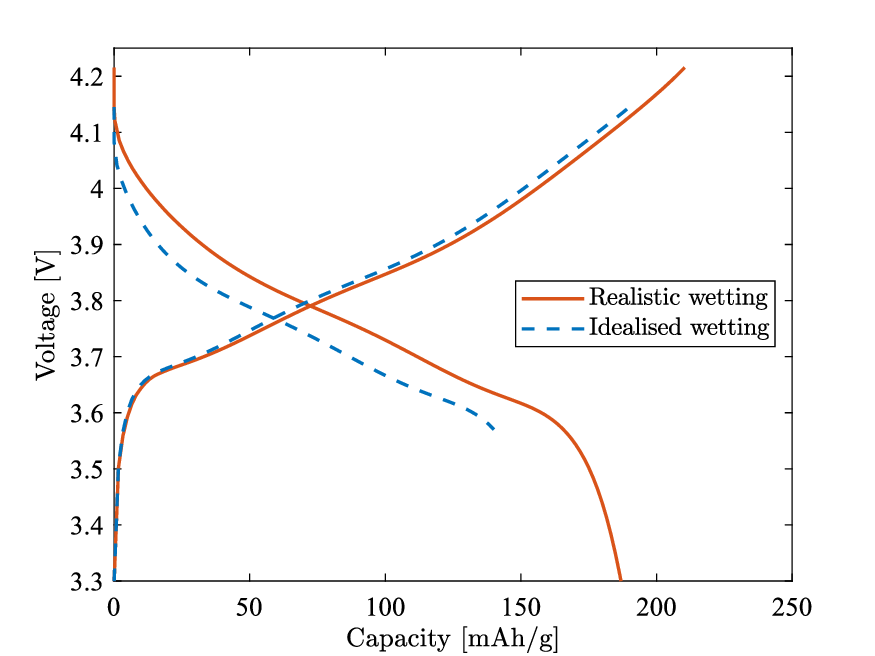}}
    \\
    \subfloat[]{\includegraphics[width=0.47\linewidth]{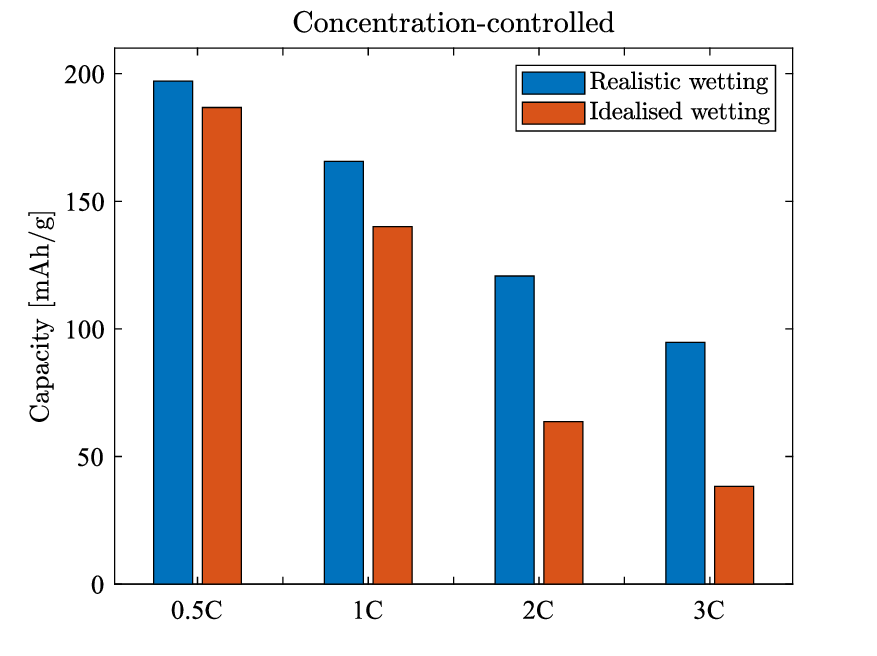}}
    \subfloat[]{\includegraphics[width=0.47\linewidth]{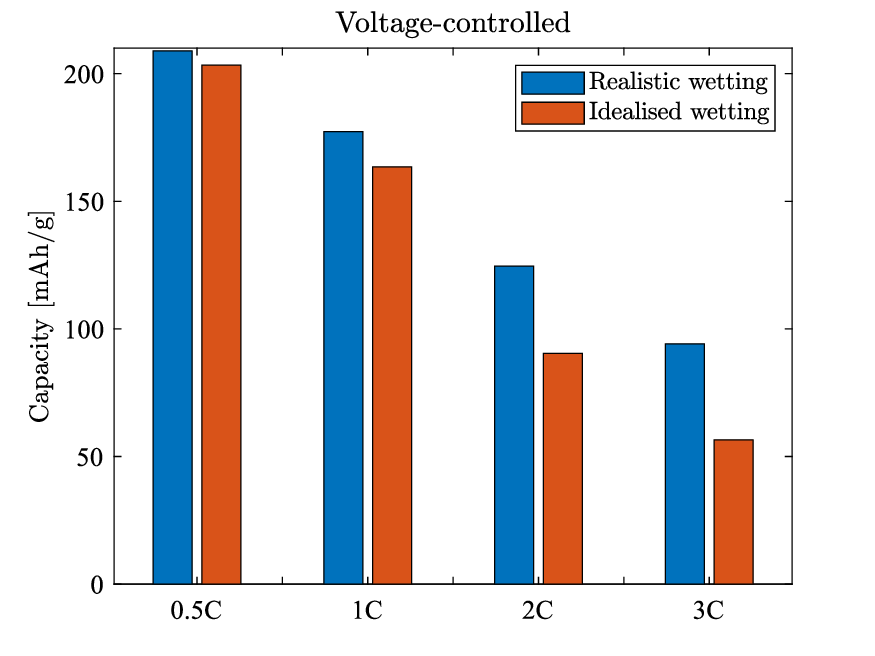}}
    \caption{Surface flux and performance impact. (a): Flux of lithium released through the cracked (blue boundary) and uncracked (orange boundary) surface, normalised by the imposed flux in the idealised wetting case $j_{app}$. (b): Potential–capacity curves for the realistic and idealised wetting cases under concentration-controlled terminal conditions. Imposing a uniform flux causes the particle to reach local concentration limits earlier, reducing the voltage window and lowering delivered capacity. (c): Delivered capacity versus C-rate for realistic wetting and idealised wetting models under concentration-controlled conditions. At 3C, the realistic case delivers over 55\% higher capacity than the idealised wetting cases. (d): Delivered capacity versus C-rate for realistic wetting and idealised wetting models under voltage-controlled conditions. At 3C, the realistic case delivers over 32\% higher capacity than the idealised wetting cases.}
    \label{fig: results 4}
\end{figure}

Figure \ref{fig: results 4}(a) shows the evolution of the (normalised) average flux $\overline{j}/j_{app}$ along the cracked and uncracked surfaces over time, normalised by the diffusion time $t_0 = R^2/D_s$. At the beginning of delithiation, the flux is uniform and equal to the reference value along the entire particle boundary. As delithiation progresses, the large flux at the crack tip induces spatial heterogeneity, leading to an increment in average lithium flux through the cracked surface and a corresponding decrease through the uncracked surface.

The impact of localised interfacial reaction redistribution on global electrochemical performance is quantified through the voltage–capacity curves shown in Fig. \ref{fig: results 4}(b). The apparent capacity delivered by a single-particle model depends on the operational criterion used to reverse current and terminate the cycle. In this work, two complementary terminal conditions are considered: a concentration-controlled criterion, representing a conservative local condition, and a voltage-controlled criterion, in which current reversal is imposed at a fixed upper voltage (4.2 V), representing global cell-level operation. 

Under concentration-controlled terminal conditions, the realistic wetting model exhibits extended delithiation and lithiation intervals compared to the idealised reference case, despite identical local concentration limits being imposed, as shown in Fig. \ref{fig: results 4}(b). This behaviour arises from the non-uniform redistribution of interfacial reaction rates, which delays the attainment of critical local lithium depletion in the realistic wetting framework. As a result, the realistic model delivers a discharge capacity of 186.9 mAh g$^{-1}$, compared with 140 mAh g$^{-1}$ for the idealised wetting case, corresponding to an apparent capacity difference of approximately 25\%. When capacities are compared at a fixed voltage, the idealised wetting case delivers 15\% less capacity than the realistic wetting model at 3.6 V. Differences in the timing at which concentration limits are reached also influence Coulombic efficiency: the realistic wetting model extracts more lithium before reaching the maximum concentration during discharge, resulting in a Coulombic efficiency of 89\%, compared with 75\% for the idealised case. Thus, under concentration-controlled operation, the primary impact of the uniform-flux assumption is a shift in the onset of local concentration limits rather than an intrinsic reduction in extractable lithium.

During the initial stages of the cycle, both models follow similar voltage-capacity trajectories. However, the idealised wetting case reaches the minimum local lithium concentration earlier, thereby initiating lithiation sooner than the realistic wetting model. This observation suggests that allowing the idealised wetting case to delithiate beyond the imposed local concentration cutoff could reduce the apparent capacity gap. This hypothesis is confirmed by adopting voltage-controlled terminal conditions, under which the particle is allowed to delithiate beyond $0.1c_{s_{max}}$. Under this criterion, the idealised case extends both delithiation and lithiation, reaching a delivered capacity of 163.5 mAh g$^{-1}$, which reduces the capacity difference relative to the realistic wetting model to approximately 12\%.

The sensitivity to the terminal criterion increases with C-rate (Fig. \ref{fig: results 4}(c)). At low C-rates, lithium can diffuse between the particle interior and surface sufficiently rapidly to maintain a more uniform concentration field, resulting in comparable delivered capacities for realistic and idealised cases. At higher C-rates, solid-state diffusion cannot keep pace with the interfacial flux, leading to unutilised active material and earlier attainment of local concentration limits, particularly under concentration-controlled operation. Comparing delivered capacity at equal final voltages between the realistic wetting model and the concentration-controlled idealised wetting case at 0.5C, 1C, 2C, and 3C, the latter delivers 5\%, 15\%, 43\%, and 55\% less capacity than the former, respectively. Figure \ref{fig: results 4}(d) shows the delivered capacity for voltage-controlled conditions. Although differences between the realistic and idealised wetting models are smaller than in the case of concentration-controlled conditions, the same overall trend is observed, with the realistic model delivering up to 32\% more capacity than the idealised wetting model at 3C.

The lower capacity observed under concentration-controlled operation is primarily a consequence of the local termination criterion. However, achieving additional capacity under voltage-controlled operation requires driving portions of the cathode particle into a low-lithium regime, in which Ni-rich layered oxides are known to undergo surface reconstruction and irreversible structural changes at high states of charge \cite{Ryu_2018_fading, Kondrakov_2017}. These transformations can lead to the formation of rock-salt surface layers that hinder lithium diffusion and ultimately promote long-term capacity fading \cite{Gan_2023}.

\subsection{Impact on particle cracking}
\label{sec: cracking}

\begin{figure}[H]
    \centering
    \subfloat[]{\includegraphics[width=0.5\linewidth]{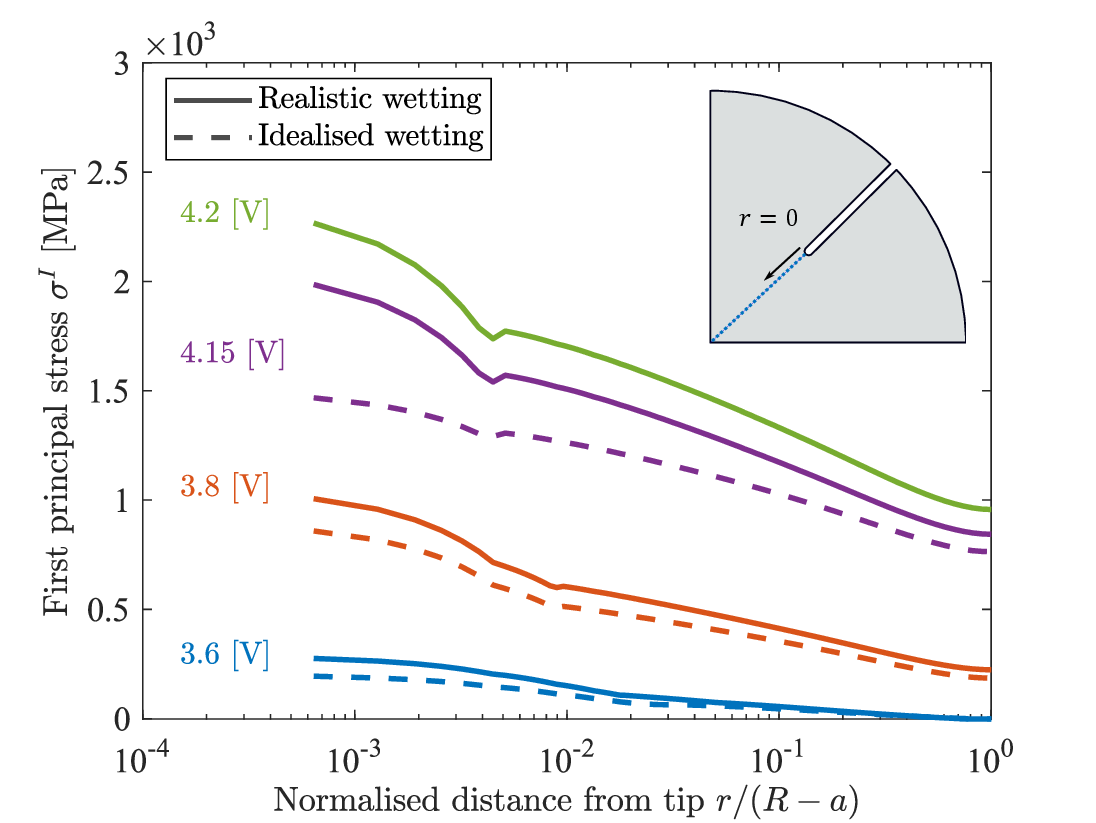}}
    \subfloat[]{\includegraphics[width=0.5\linewidth]{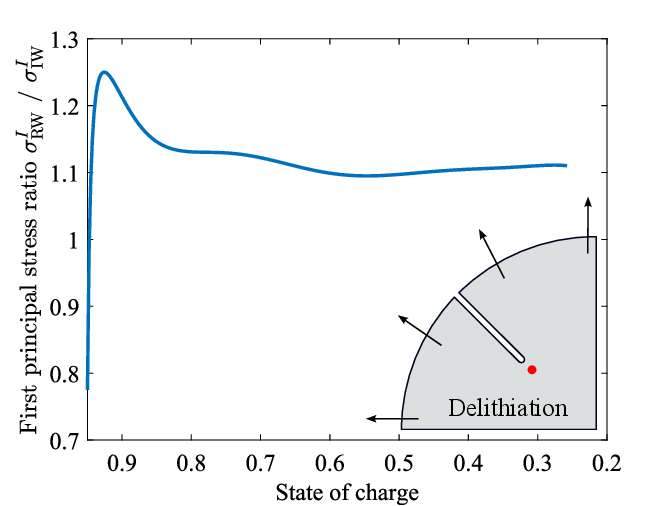}}
    \caption{Stresses during delithiation process. (a): First principal stress profiles from the crack tip toward the particle centre at different voltages during delithiation. Solid lines correspond to the realistic wetting model and dashed lines to the idealised wetting case. (b): Ratio of first principal stress in the realistic wetting (RW) model to that in the idealised wetting (IW) case, evaluated at $r=0.1\,(R-a)$ from the crack tip (with $R$ the particle radius and $a$ the crack length) during delithiation. Over the full delithiation process, the realistic wetting model predicts stresses at least 10\% higher than the idealised case.}
    \label{fig: results 5}
\end{figure}

To assess the impact of wetting on crack evolution, we begin by analysing the stress response of the cathode particle during delithiation. Figure \ref{fig: results 5}(a) compares the first principal stress predicted by the realistic and idealised wetting models. While both cases exhibit comparable stress levels over most of the delithiation process, the underlying mechanisms differ due to the redistribution of interfacial reactions in the realistic wetting model.

In the realistic model, the lithium flux at the particle boundary is coupled to both the local concentration and the stress state. Regions under tensile hydrostatic stress promote lithium accumulation through the stress-dependent chemical potential (Eq. (\ref{eq: chemical potential})), while local variations in concentration influence the overpotential via the equilibrium potential (Eq. (\ref{eq: overpotential})), enhancing lithium extraction. This establishes a competition between stress-driven accumulation and reaction-driven depletion, which regulates the local concentration field and leads to a more uniform lithium distribution within the particle and along its surface.

In contrast, the idealised framework imposes a uniform interfacial flux that is independent of local concentration, preventing this feedback mechanism. As a result, higher concentration gradients develop, causing delithiation to terminate at a lower voltage in the idealised case (4.15 V versus 4.2 V). The extended delithiation window in the realistic framework consequently leads to higher stress levels than in the idealised case.

The electrolyte-enabled reaction redistribution introduces a self-regulating mechanism that limits local lithium accumulation, which is absent in uniform flux models. As a result, only the accumulation mechanism is active in the idealised framework, leading to higher lithium concentrations near the crack tip compared to the realistic case. This lower concentration generates larger chemical strains (Eq. (\ref{eq: concentration strain})) and, consequently, higher stresses in the realistic wetting model, as shown in Fig. \ref{fig: results 5}(b). At equivalent states of charge, the idealised case predicts lower stress levels, with differences exceeding 10\% throughout delithiation.

\begin{figure}[H]
    \centering
    \subfloat[]{\includegraphics[width=0.47\linewidth]{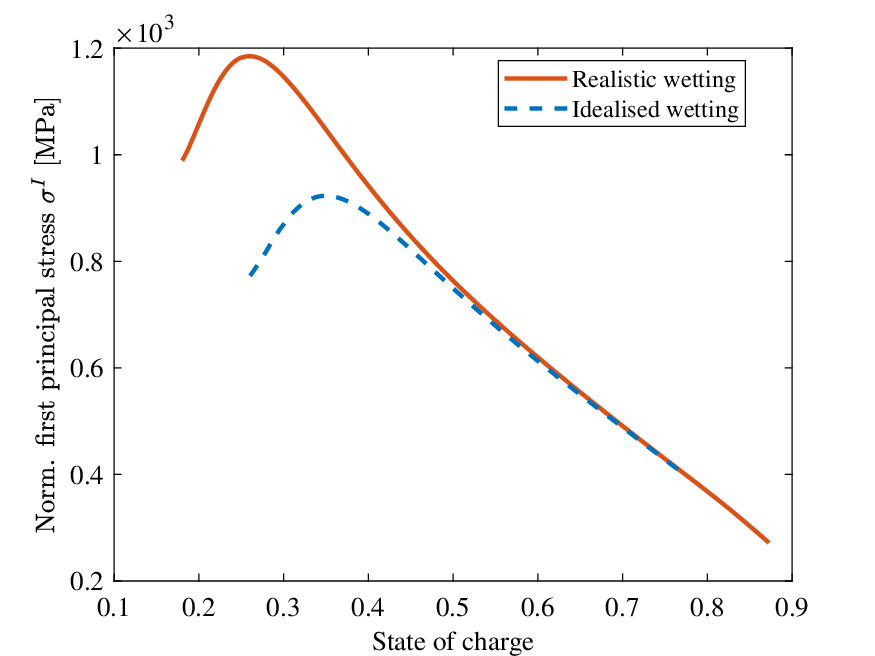}}
    \subfloat[]{\includegraphics[width=0.47\linewidth]{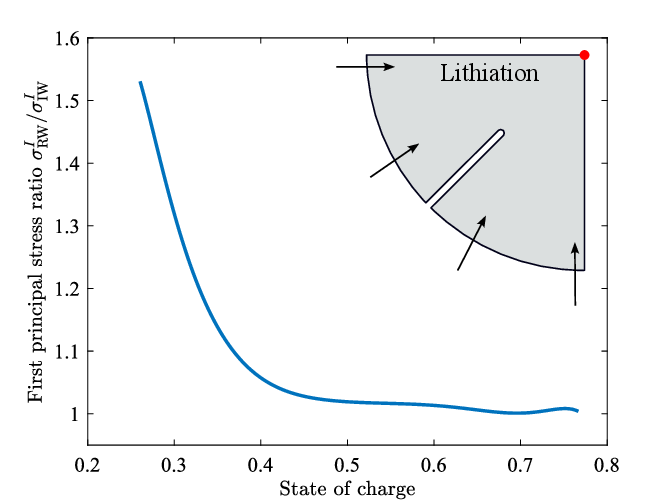}}
    \\
    \subfloat[]{\includegraphics[width=0.47\linewidth]{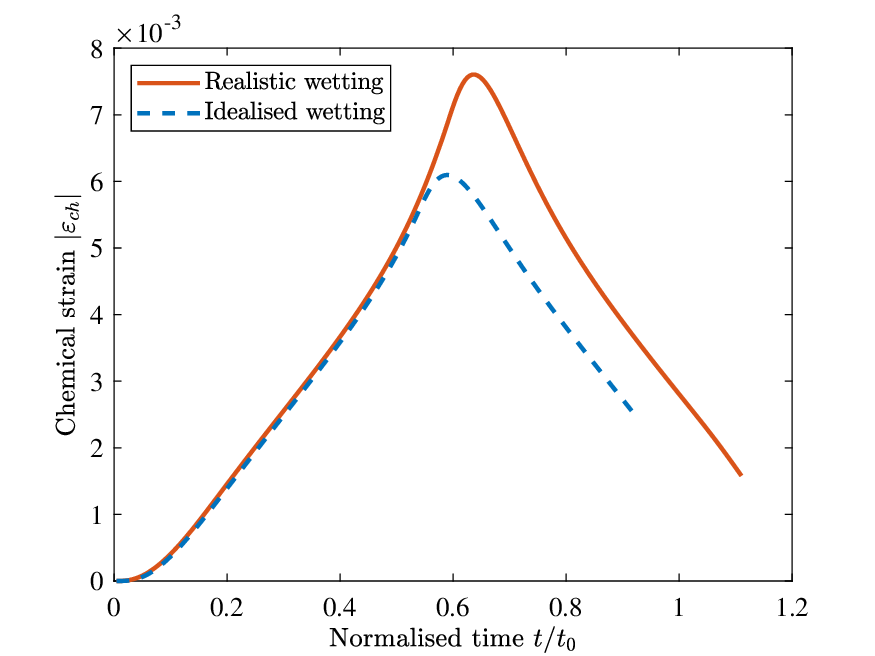}}
    \subfloat[]{\includegraphics[width=0.47\linewidth]{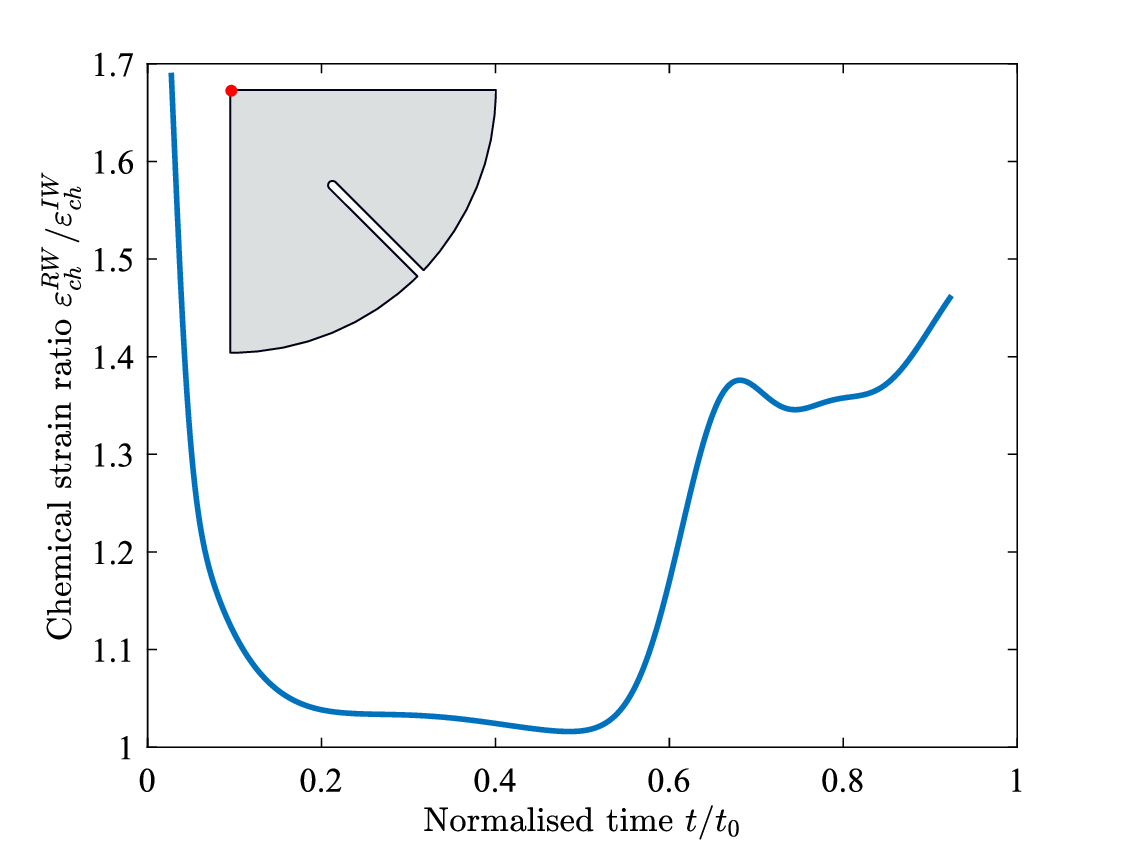}}
    \caption{Stress and strain evolution at the particle centre. (a): First principal stress at the particle centre for the realistic wetting and idealised wetting cases during lithiation. (b): Ratio of first principal stress in the realistic wetting (RW) model to that in the idealised wetting (IW) case at the particle centre during lithiation. (c): Chemical strain at the particle centre as a function of normalised time $t/t_0$, where $t_0 = R^2/D_s$ is the characteristic diffusion time. (d) Ratio of chemical strain in the RW model to that in the IW case, evaluated at the particle centre, as a function of $t/t_0$. The largest differences in chemical strain occur near the delithiation-lithiation transition.}
    \label{fig: results 6}
\end{figure}

During lithium insertion, the largest tensile stress happens at the particle centre; for this reason, in Fig. \ref{fig: results 6}(a), we plot the first principal stresses for the realistic and idealised case at the particle centre during lithiation. Because the idealised wetting case experiences lower stresses at the end of delithiation, the initial stress at the onset of lithiation is also lower at the particle centre. At the beginning of discharge, the higher lithium concentration at the particle centre relative to the boundary drives lithium diffusion toward the surface. As lithium insertion continues, the concentration increases throughout the particle and particularly at the surface, leading to a reduction and eventually change of direction of the concentration gradient. These changes in the concentration evolution explain the initial increase and subsequent decrease in stress at the particle centre.

Since lithiation begins at different states of charge in the two cases, the stress at the particle centre in the realistic wetting model starts to decrease while the stress in the idealised wetting case is still increasing. This mismatch in trends produces a steep decrease in the ratio of first principal stress between the realistic and idealised cases, as shown in Fig. \ref{fig: results 6}(b). After a state of charge of approximately 0.4, the particle-centre stresses become nearly identical in both cases, indicating that interfacial reaction redistribution primarily affects stress evolution during the delithiation-lithiation transition.

These differences are also reflected in the evolution of chemical strain $\varepsilon_{ch}$ as a function of time (normalised by the characteristic diffusion time $t_0 = R^2/D_s$), as shown in Figs. \ref{fig: results 6}(c, d). Similar to the stress response, the chemical strain at the particle centre increases nearly identically in both cases up to $t/t_0 = 0.5$. Beyond this point, the realistic wetting model exhibits a more pronounced and prolonged deformation. This behaviour arises from the lower lithium concentration reached at the particle centre due to the extended delithiation period in the realistic wetting model compared with the idealised case.

These results demonstrate that imposing a uniform flux at the particle boundary leads to a systematic underestimation of the stress levels, with implications for particle cracking predictions. For example, defining the local stress intensity factor as $K_I \approx \sigma^I \sqrt{\pi a}$, with $\sigma^I$ being the first principal stress and $a$ the crack length, results in $K_I \sim 1~\mathrm{MPa~m^{1/2}}$ for the electrolyte wetting case at 3.9 V. This would be sufficient to trigger cracking, since fracture toughness of Ni-rich cathode materials is approximately $K_{IC} \approx 1~\mathrm{MPa~m^{1/2}}$ \cite{pandurangi_chemo-mechanical_2023}, but this would not be the case for the constant flux/idealised wetting, where $K_I$ is 10\% lower under those conditions.

The reaction redistribution mechanism unveiled by the present work is likely to be more pronounced for particle geometries and crack configurations that favour strong internal concentration gradients (e.g., large particles containing long and sharp cracks). Cathode electrolyte interphase (CEI) formation and growth, not considered in this study, could also influence the findings quantitatively, as it would introduce additional interfacial resistance and alter local charge-transfer kinetics. This would be particularly the case on newly exposed crack surfaces at high voltage, where the reaction amplification at the crack tip could be diminished. However, all cases considered here operate at or below 4.2 V, under which electrolyte oxidation is expected to be less severe \cite{xu2014electrolytes}.

\section{Conclusions}
\label{Sec:Concluding remarks}

In this work, we investigated the effect of interfacial reaction redistribution in cracked cathode particles by comparing a realistic wetting electro-chemo-mechanical model, which explicitly resolves electrolyte behaviour inside and outside cracks, with an idealised wetting chemo-mechanical model that assumes a uniform lithium flux at the particle surface. This comparison isolates the role of electrolyte coupling in lithium transport, stress evolution, and extractable capacity during a full charge-discharge cycle. Crack–electrolyte coupling is found to redistribute interfacial reactions, modifying the resulting concentration and stress histories. Our results reveal that:
\begin{itemize}
    \item Due to the narrower and more confined transport pathways inside cracks, the electrolyte potential at the crack tip is approximately 45\% higher than in the bulk electrolyte, although these variations remain small relative to the dominant solid-state driving forces.

    \item The realistic wetting model reveals strong spatial heterogeneity in interfacial reaction rates during delithiation, leading to local lithium flux amplification at the crack tip, reaching almost 8 times the value predicted by the idealised case at 1C-rate.

    \item Under concentration-controlled conditions, the idealised model underpredicts delivered capacity by 25\% relative to the realistic wetting case, and yields a lower Coulombic efficiency (75\% versus 89\%), due to earlier attainment of local concentration limits.

    \item Under global operation conditions (voltage-controlled), capacity differences are reduced but remain sensitive to the C-rate. At 3C, the idealised wetting model underpredicts delivered capacity by up to 32\% at equal terminal voltage.
\end{itemize} 

Overall, uniform flux single-particle models underestimate both utilisation limits and stress levels near the delithiation-lithiation transition. While electrolyte wetting primarily acts as a geometric pathway that enables additional electrochemical reactions, the resulting redistribution is governed mainly by variations in local concentrations in the active material and stress fields. Thus, neglecting crack-electrolyte coupling leads to quantitative underestimation of performance and fatigue-relevant stress histories in cracked cathode particles.

\section*{Acknowledgments}
\label{Acknowledge of funding}

\noindent S. Luza-Vega acknowledges financial support from the National Agency for Research and Development (ANID), Chile/Scholarship Program/DOCTORADO BECAS CHILE/2023 - 72230336. E. Mart\'{\i}nez-Pa\~neda acknowledges financial support from UKRI's Future Leaders Fellowship programme [grant MR/V024124/1]. Y. Zhao is supported by the National Natural Science Foundation of China (Project No. 12472175).




\end{document}